\documentclass[10pt, twocolumn]{IEEEtran}

\usepackage{cite}
\usepackage{graphicx,enumerate} 
\graphicspath{./imagesJsac_2018}

\usepackage{amsfonts}
\usepackage{mathtools}
\usepackage{amsmath}
\usepackage{color}

\usepackage{stackrel}

\newcommand{\Lim}[2]{\displaystyle \lim_{{#1}\rightarrow{#2}} \hspace{-1mm}}
\newtheorem{pb}{Problem}
\newtheorem{res}{Result}
\newtheorem{lem}{Lemma}

\usepackage{array}

\usepackage{algorithm}
\usepackage{algorithmicx}
\usepackage{algpseudocode}

\ifCLASSOPTIONcompsoc
    \usepackage[caption=false, font=normalsize, labelfont=sf, textfont=sf]{subfig}
\else
	\usepackage[caption=false, font=footnotesize]{subfig}
\fi

\begin{document}

\title{Energy-Latency Tradeoff in Ultra-Reliable Low-Latency Communication with Retransmissions}

\author{\IEEEauthorblockN{Apostolos~Avranas\IEEEauthorrefmark{1}, Marios~Kountouris\IEEEauthorrefmark{1}, and Philippe Ciblat\IEEEauthorrefmark{2}} %
	
	\IEEEauthorblockA{\IEEEauthorrefmark{1}Mathematical and Algorithmic Sciences Lab, Paris Research Center, Huawei France}\\
	\IEEEauthorblockA{\IEEEauthorrefmark{2}T\'el\'ecom ParisTech, Universit\'e Paris-Saclay, F-75013 Paris, France} %
	
	Emails: \{apostolos.avranas,marios.kountouris\}@huawei.com, philippe.ciblat@telecom-paristech.fr
}

\maketitle

\begin{abstract}
High-fidelity, real-time interactive applications are envisioned with the emergence of the Internet of Things (IoT) and tactile Internet by means of ultra-reliable low-latency communications (URLLC). Exploiting time diversity for fulfilling the URLLC requirements in an energy efficient manner is a challenging task due to the nontrivial interplay among packet size, retransmission rounds and delay, and transmit power. In this paper, we study the fundamental energy-latency tradeoff in URLLC systems employing incremental redundancy (IR) hybrid automatic repeat request (HARQ). We cast the average energy minimization problem with a finite blocklength (latency) constraint and feedback delay, which is non-convex. We propose a dynamic programming algorithm for energy efficient IR-HARQ optimization in terms of number of retransmissions, blocklength and power per round. Numerical results show that our IR-HARQ approach could provide around 25\% energy saving compared to one-shot transmission (no HARQ).
\end{abstract}

\begin{IEEEkeywords}
	URLLC, tactile Internet, IR-HARQ, energy minimization, finite blocklength.
\end{IEEEkeywords}

\section{Introduction}
Current wireless networks have typically been designed for increasing throughput and improving coverage, focusing mainly on human-centric communication and delay-tolerant content. The emergence of the Internet of Things (IoT) we experience nowadays enables a transition towards device-centric communication and real-time interactive systems. Various socially useful applications and new uses of wireless communication are currently envisioned in areas such as industrial control, smart cities, augmented/virtual reality (AR/VR), automated transportation, and tactile Internet. Tactile Internet enables real-time connection between people and objects and will be instrumental for supporting low-latency, high-fidelity, control-type applications, such as telesurgery, remote driving, and industrial remote monitoring \cite{Tactile1,Tactile2}.
The mission critical and societal aspect of tactile Internet makes the support for very low end-to-end latency and extreme reliability required. The tolerable latency for tactile Internet has been set to 1 ms and ultra-reliability is quantified in terms of outage probability of $10^{-5}$ or even $10^{-7}$. Ultra-reliable low-latency communications (URLLC) lies in the overlapped area of the IoT and tactile Internet and is a key technology pillar in emerging mobile networks. Fifth generation (5G) systems envision to support URLLC scenarios with strict requirements in terms of latency (ranging from 1 ms and below to few milliseconds) and reliability (higher than 99.999\%). 

Guaranteeing the URLLC requirements is a challenging task since the performance is constrained by fundamental tradeoffs between delay, throughput, energy and error probability. The predominance of short messages for mission critical IoT, together with the need to reduce the packet duration and channel uses, impose that small blocklength channel codes are also used. This results in a rate penalty term and transmission rates with non-zero error probability, revisiting key insights obtained via asymptotic information theoretic results. Recent progress has quantified the effect of finite blocklength, providing tight bounds and accurate normal approximation for the maximum coding rate to sustain the desired packet error probability for a given packet size \cite{PolyanskiyThesis,Hayashi}. In order to compensate for the reliability loss introduced by short packets, reliable communication mechanisms creating diversity have to be carried. A standard technique to improve transmission reliability, which has been adopted in various wireless standards, is incremental redundancy (IR) hybrid automatic repeat request (HARQ). However the benefits of time diversity could be rather limited under stringent latency constraints as the number of transmission rounds and channel uses is rather limited. Moreover, the benefit of feedback-based retransmissions (even with error-free but delayed feedback) is questionable since each transmit packet is much smaller due to energy and latency constraints, thus more prone to errors. Additionally, energy considerations, in particular power consumption, are of cardinal importance in the design of tactile Internet, and there is an inherent energy-latency tradeoff. A transmission can be successful with minimum delay at the expense of additional or high power usage. In the short-packet regime, this interplay is more pronounced as latency is minimized when all packets are jointly encoded, whereas power is minimized when each packet is encoded separately. The general objective of this work is to characterize the fundamental energy-latency tradeoff and optimize IR-HARQ in URLLC systems. 

\subsection{Related work}
Prior work has considered the problem of throughput maximization by either adjusting the blocklength of each IR-HARQ round using the same power \cite{MakkiFinite} or via rate refinement over retransmissions of equal-sized and constant energy packets \cite{JindalHARQ}. Equal-sized and constant energy packets and rate maximization under a reliability constraint is considered in \cite{MaxRateGivenErrorPowerLength}. In \cite{SpherePackingHARQ}, sphere packing is used for optimizing the blocklength of every transmission with equal power. Dynamic programming for optimizing HARQ protocols with long packets is done in \cite{RatePowerHARQ,OptARQdynamicFirst,SzczecinskiRateAllocDynamic}. In \cite{SchedulingConvexityFinite}, both power and blocklength are tuned so as to minimize the energy consumed by packets scheduled in a FIFO manner. 
Finally, \cite{PopovskiEMS} proposes a new family of protocols and compares its throughput with a dynamically optimized IR-HARQ.

\subsection{Contributions}
In this paper, we analyze the fundamental tradeoff between latency (in terms of feedback and retransmission delay) and average consumed energy in the finite blocklength regime for URLLC systems with IR-HARQ. Considering that packets have to be decoded with a certain error probability and latency, we provide an answer whether it is beneficial to do one-shot transmission (no HARQ) or split the packet into sub-codewords and use IR-HARQ. We propose a dynamic programming algorithm for energy efficient IR-HARQ optimization in terms of number of retransmissions, blocklength and power per round. Furthermore, the impact of feedback delay on the energy consumption and IR-HARQ performance is also investigated. Finally, we investigate the asymptotic (infinite blocklength) regime and derive an expression for the solution of the average energy minimization problem. Numerical results show that our IR-HARQ approach could provide around 25\% energy saving compared to one-shot transmission.


\section{System Model}\label{sec:model}
We consider a point-to-point communication link, where the transmitter has to send $B$ information bits within a certain predefined latency, which can be expressed by a certain predefined maximum number of channel uses, denoted by $N$. If no ARQ/HARQ mechanism is utilized, the packet of $B$ bits is transmitted only once (one-shot transmission) and its maximum length is $N$. When a retransmission strategy is employed, we consider hereafter IR-HARQ with $M$ transmissions (rounds), i.e., $M-1$ retransmissions. Setting $M=1$, we recover the no-HARQ case as a special case of the retransmission scheme. We denote $n_m$ with $m\in\lbrace 1, 2, ..., M\rbrace$ the number of channel uses for the $m$-th transmission. 

The IR-HARQ mechanism operates as follows: $B$ information bits are encoded into a parent codeword of length $\sum_{m=1}^M n_m$ symbols. Then, the parent codeword is split into $M$ fragments of codeword (sub-codewords), each of length $n_m$. The receiver requests transmission of the $m$-th sub-codeword only if it is unable to correctly decode the message using the previous 1 to $m-1$ fragments of the codeword. In that case, the receiver concatenates the first till $m$-th fragments and attempts to jointly decode it. We assume that the receiver knows perfectly whether or not the message is correctly decoded (through CRC) and ACK/NACK is received error free but with delay. Every channel use (equivalently the symbol) requires a certain amount of time, therefore we measure time by the number of symbols contained in a time interval. The latency constraint is accounted for by translating it into a number of channel uses as follows: we have $\sum_{m=1}^M \left(n_m+D(\vec{n}_m) \right)\leq N$ where $\vec{n}_m$ is the tuple $(n_1,n_2,..., n_m)$ and $D(\cdot)$ is a penalty term introduced at the $m$-th transmission due to delay for the receiver to process/decode the $m$-th packet and send back acknowledgment (ACK/NACK). The penalty $D(\cdot)$ on the $m$-th round may depend on the previous transmissions, i.e., $\vec{n}_{m-1}$, since IR-HARQ is employed and the receiver applies a decoding processing over the entire $\vec{n}_m$.

The channel is considered to be quasi-static along with the whole HARQ mechanism, i.e., the channel coefficients remain constants during the packet retransmissions. This is a relevant model for short-length packet communication and IoT applications. 
For a system operating at carrier frequency $f_c = 2.5$ GHz and coherence time $T_c=1$ ms (latency constraint), the receiver speed is $v = cB_d/f_c \approx 180$ km/h, where $B_d=0.423/T_c$ \cite[(8.20)]{BookCommGibson}, is the Doppler spread $c$ is the speed of light; this is a relatively high speed for most mission-critical IoT or tactile Internet applications. Therefore, our communication scenario consists of a point-to-point link with additive white Gaussian noise (AWGN). Specifically, in $m$-th round, the fragment (sub-codeword) $c_m\in \mathbb{C}^{n_m}$ is received with power $P_m=\frac{||c_m||^2}{n_m}$ and distorted by an additive white circularly-symmetric complex Gaussian random process with zero mean and unit variance. As the channel is static along with the transmission, the channel gains are constant and the noise variance is assumed equal to one without loss of generality. The power allocation applied during the the first $m$ rounds is denoted by $\vec{P}_m=(P_1,... P_m)$. 
   
\section{Problem Statement and Preliminaries}\label{sec:Preliminaries}
The objective of this paper is twofold: i) to derive the best HARQ mechanism that minimizes the average consumed energy for a given packet error probability and latency constraint (URLLC requirements) by optimally tuning both $\vec{n}_M$  and $\vec{P}_m$ for a prefixed $M$ (number of transmissions per HARQ mechanism), and ii) to find the optimal number of transmission rounds $M$ for different feedback delay models. 

The first step for reaching the above objectives is to characterize the probability of error in the $m$-th round of the HARQ mechanism as a function of $\vec{n}_m$ and $\vec{P}_m$. To derive the packet error probability in short-packet communication, we resort to results for the non-asymptotic (finite-blocklength) regime \cite{PolyanskiyThesis}. 

In IR-HARQ with $(m-1)$ retransmissions, the packet error probability or equivalently the outage probability, denoted by $\epsilon_m$, can be expressed as $\displaystyle \epsilon_m=\mathbb{P}(\bigcap_{i=1}^m\Omega_i)$ where $\Omega_m$ is the event ''the concatenation of the first $m$ fragments of the parent codeword, which have length $\vec{n}_m$ and energy per symbol $\vec{P}_m$, is not correctly decoded assuming optimal coding''. 

When an \textit{infinitely} large blocklength is assumed, an error occurs if the mutual information is below a threshold and for IR-HARQ, it can easily be seen that for $k<m$ we have $\Omega_m\subseteq \Omega_k$ \cite{caire-tuninetti,leduc}, which leads to $\epsilon_m=\mathbb{P}(\Omega_m)$. In contrast, when a real coding scheme (and so \textit{finite} blocklength)  is used, the above statement does not hold anymore and an exact expression for $\epsilon_m$ seems intractable. Therefore, in the majority of prior work on HARQ (see \cite{leduc} and references therein), the exact outage probability $\epsilon_m$ is replaced with the simplified $\varepsilon_m$ defined as $\displaystyle \varepsilon_m=\mathbb{P}(\Omega_m)$, since $\varepsilon_m$ and $\epsilon_m$ perform quite closely when evaluated numerically. Note that for $m=1$ the definitions coincide and $\varepsilon_1=\epsilon_1=\mathbb{P}(\Omega_1)$). In the remainder of the paper, we assume that this approximation is also valid in the finite blocklength regime \cite{PolyanskiyThesis}. Then, $\varepsilon_m$ can be upper bounded \cite[Lemma 14]{PolyanskiyThesis} and also lower bounded as in \cite{ParksParallelAWGN} by  employing the $\kappa\beta$-bounds proposed in \cite{PolyanskiyThesis}. Both bounds have the same first two dominant terms and the error probability is approximately given by 
\begin{equation}\label{eq:epsb}
\varepsilon_m \approx Q\left(\frac{\displaystyle\sum_{i=1}^m n_i\ln(1+P_i)-B\ln2}{\sqrt{\displaystyle\sum_{i=1}^m\frac{n_iP_i(P_i+2)}{(P_i+1)^2}}}\right )
\end{equation}
where $Q(x)$ is the complementary Gaussian cumulative distribution function. For the sake of clarity, we may show the dependency on the variables, i.e., $\varepsilon_m(\vec{n}_m,\vec{P}_m)$ or $\varepsilon_m(n_1,n_2,...,P_1,P_2,...)$ instead of $\varepsilon_m$, whenever needed. 
	
\section{Optimization Problem}\label{sec:optim}
We employ an IR-HARQ with $M-1$ retransmissions with variable blocklengths and powers over rounds. We first address the problem of minimizing the average energy consumed to achieve a target reliability $T_{\rm rel}$ (e.g. $T_{\rm rel}=99.999\%$ in 3GPP URLLC or equivalently an outage probability $P_{\rm out}=1-T_{\rm rel}=10^{-5}$) without violating the latency constraint $\sum_{m=1}^M \left(n_m+D(\vec{n}_m) \right)\leq N$ by properly setting $\vec{n}_M$ and $\vec{P}_M$. 

\subsection{Optimization Problem} 
Letting $\varepsilon_0=1$, the problem is mathematically formulated as follows:

\begin{pb}\label{prob} Minimization of the average energy consumed by a HARQ mechanism leads to 
\begin{align}
\underset{\vec{n}_M,\vec{P}_M}{\min} \quad &\sum_{m=1}^M n_mP_m\varepsilon_{m-1} \label{eq:cost}\\
\quad\text{s.t.} \quad& \sum_{m=1}^M \left(n_m+D(\vec{n}_m) \right)\leq N\label{eq:C1}\\
 & \varepsilon_M\leq 1-T_{\rm rel}\label{eq:C2}\\
 & \vec{n}_M \in \mathbb{N}_{+,*}^{M}\\
 & \vec{P}_M \in \mathbb{R}_{+}^M
\end{align}
where $\mathbb{N}_{+,*}$ is the set of positive integers, and $\mathbb{R}_{+}$ corresponds to the set of non-negative real-valued variables.
\end{pb}
\smallskip
We consider two different models for the feedback delay:
\begin{itemize}
\item The first model assumes a constant delay per retransmission, i.e., $D(\vec{n}_m)=d$. This simple model corresponds to the current real communication systems (e.g. 3GPP LTE) where the feedback is sent back through frames that are regularly spaced in time. 
\item The second model assumes a non-constant delay per retransmission and that feedback is sent right after the decoding outcome at the receiver side. In that case, the limiting factor to send back the feedback is the processing time required by the receiver to decode the message. We consider this time to be proportional to the size of the set of sub-codewords the receiver has already received. Therefore, after the $m$-th transmission, we have $D(\vec{n}_m)=r\sum_{i=1}^m n_i$ with $r$ a predefined constant.\end{itemize}

In real-world systems, due to protocol constraints and implementation complexity, the same number of symbols per transmission is used ($n_m=n$, $\forall m$) and one can optimize only the power per transmission. In this work, we address the general case of variable blocklength per transmission as a means to study the maximum capability of IR-HARQ to improve the performance. Evidently, having fixed block size per transmission is a simplified version of our general problem. 

\begin{figure}[htb]
\centering
\includegraphics[width=0.9\columnwidth]{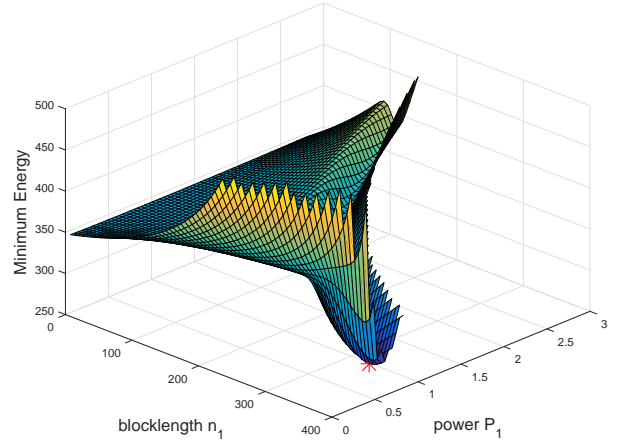}
\caption{Average consumed energy versus $(n_1,P_1)$ for $N=400$, $B=32$ bytes, and $T_{\rm rel}=99.999\%$. The red asterisk marks the minimum.}
\label{Fig:Non-Convexity}
\end{figure}

Problem \ref{prob} is a Mixed Integer Nonlinear Programming (MINLP) problem and a first approach to overcome its hardness is to relax the integer constraint by looking for $\vec{n}_M\in\mathbb{R}_{+,*}^M$ instead of $\vec{n}_M\in\mathbb{N}_{+,*}^M$. Even with that relaxation, the problem remains hard in the sense that the non-linearity cannot be managed through convexity properties of the relaxed problem. Indeed, in Figure~\ref{Fig:Non-Convexity} we plot the objective function of Problem \ref{prob} for $M=2$, $D(\vec{n}_m)=0$ and equality in the latency and reliability constraints, i.e., \eqref{eq:C1} and \eqref{eq:C2} in order to have only a 2D search on variables $(n_1,P_1)$. We observe that the objective function is neither convex nor quasi-convex nor biconvex, consequently standard convex optimization methods cannot be used.
 
Therefore, our objective is not providing a closed-form optimal solution for Problem \ref{prob} but deriving a low complexity algorithm finding the optimal solution. In the next two subsections, we show that Problem \ref{prob} can be written with equality in its constraints, and that a dynamic programming algorithm can be used to find the optimal solution.

\subsection{Low Complexity Algorithm with Equality Constraints}
We first start with the simple case where no delay penalty is considered ($D(\vec{n}_m)=0, \forall m$). 

\smallskip
\begin{res}\label{res:equality}
When $D(\vec{n}_m)=0, \forall m$, the optimal solution  of Problem \ref{prob} satisfies the latency constraint given by \eqref{eq:C1} and the reliability constraint given by \eqref{eq:C2} with equality. 
\end{res}
\smallskip
This result has two consequences: 
\begin{enumerate}[(i)]
\item Equality in \eqref{eq:C1} and \eqref{eq:C2} enables us to reduce the number of variables since one $n_m$ and one $P_m$ can be removed from the unknown variables, i.e., we search over $2(M-1)$ instead of $2M$ variables. In the conference version of our work \cite{apos_globecom}, we have treated the case of $M=2$, which leads to a 2D search instead of a 4D search. 
But as $M$ becomes larger, the two equalities are insufficient to significantly reduce the computational cost of the optimization algorithm. 
\item Equality in \eqref{eq:C1} implies that it is advantageous to send as many symbols as possible during transmission but with less energy used for each symbol. In other words, \emph{given an energy budget, it is preferable to spread this budget into many symbols with low power rather than to few ones with high power}. 
\end{enumerate}

Proving the above result requires the following lemmas:
\begin{lem}\label{lem:1}
The optimal solution of Problem \ref{prob}, denoted by $(\vec{n}_M^\star,\vec{P}_M^\star)$, satisfies $\varepsilon_{M-1}>\varepsilon_M$.
\end{lem} 
\begin{IEEEproof}
Let $\vec{P}_M^\star=(\vec{P}_{M-1}^\star,P_M^\star)$. If $\varepsilon_{M-1}\leq\varepsilon_M$ at  $(\vec{n}_M^\star,\vec{P}_M^\star)$, then $(\vec{n}_M^\star,\vec{P}_M')$ with $\vec{P}_M'=(\vec{P}_{M-1}^\star,0)$ offers a lower consumed average energy since the last term in the sum of the objective function can be removed while the other terms remain identical. This leads to a contradiction preventing $\varepsilon_{M-1}\leq\varepsilon_M$ at the optimal point.
\end{IEEEproof}

\begin{lem}\label{res:epsbDecreasingP2} 
If $(\vec{n}_M^\dag,\vec{P}_M^\dag)$ satisfies $\varepsilon_{M-1}>\varepsilon_M$, then the function $P\mapsto \varepsilon_M(\vec{n}_M^\dag, \vec{P}_{M-1}^\dag, P)$ is decreasing in the neighborhood of $P_M^\dag$.
\end{lem}
\begin{IEEEproof}
See Appendix \ref{anx:res1}.
\end{IEEEproof}
\smallskip

Lemma~\ref{res:epsbDecreasingP2} enables us to force the constraint \eqref{eq:C2} to be satisfied in equality, and so proves the second part of Result \ref{res:equality}. To prove that, we assume that the optimal point $(\vec{n}_M^\star,\vec{P}_M^\star)$ satisfies $\varepsilon_M<1-T_{\rm rel}$. According to Lemma~\ref{lem:1}, we know that $\varepsilon_{M-1}>\varepsilon_M $. Consequently, according to Lemma~\ref{res:epsbDecreasingP2}, $P_M^{\star}$ can be decreased to $P_M'$ such that $\varepsilon_M<1-T_{\rm rel}$ is still true (due to continuity of the function). This implies that  $(\vec{n}_M^\star,\vec{P}_{M-1}^\star,P_M')$ is a better solution than the optimal one, which leads to contradiction preventing $\varepsilon_M<1-T_{\rm rel}$ at the optimal point.

For proving that equality in constraint \eqref{eq:C1} is required at the optimal point, we need to establish the following result.
\begin{lem}\label{lem:NbetterThanP}
Let $\mathcal{B}{=}\lbrace(n_1, \cdots , n_M, P_1, \cdots, P_M){\in} \mathbb{R}_{+,*}^{2M} | 0.5>\varepsilon_1(n_1,P_1)>\varepsilon_M(n_1,..,n_M,P_1,...,P_M){>} Q(\sqrt{2B\ln2}/3)\rbrace$. As long as $(an_1,n_2,..,n_M,P_1/a,P_2,...,P_M) \in \mathcal{B}$, $\varepsilon_1(an_1,P_1/a)$ and $\varepsilon_M(an_1,n_2,..,n_M,P_1/a,P_2,...,P_M)$ are decreasing with respect to $a$.
\end{lem} 
\begin{IEEEproof}
	See Appendix \ref{anx:res2}.
\end{IEEEproof}
\smallskip

Lemma~\ref{lem:NbetterThanP} enables us to force the constraint \eqref{eq:C1} to be satisfied in equality, and so proves the first part of Result \ref{res:equality} as soon as the optimal point belongs to $\mathcal{B}$, i.e., satisfies $0.5{>}\varepsilon_1{>}\varepsilon_M{=}1-T_{\rm rel}{>}Q(\sqrt{2B\ln2}/3)$. To prove that, we assume that the optimal point $(\vec{n}_M^\star,\vec{P}_M^\star)$ satisfies $\sum_{m=1}^M n_m^{\star}<N$. For any $a>1$ such that $(an_1^{\star},n_2^{\star},..,n_M^{\star},P_1^{\star}/a,P_2^{\star},...,P_M^{\star}){\in} \mathcal{B}$ and $an_1^{\star}+\sum_{m=2}^M n_m^{\star}\leq N$ yields a better solution. And there exists at least one $a>1$ in $\mathcal{B}$ by continuity of  $\varepsilon_1$ and $\varepsilon_M$ with respect to $a$. Actually $an_1
^\star$ may belong to $\mathbb{R}_{+,*}$ instead of  $\mathbb{N}_{+,*}$. To overcome this issue, we assume that the scheme with $a=(n_1^{\star}+1)/n_1^{\star}$ is still in $\mathcal{B}$, i.e., increasing the blocklength of the first fragment by one symbol does not bring us out of $\mathcal{B}$.

We consider now the case of $D\neq 0$. The nonzero feedback delay does not modify Result \ref{res:equality} for the reliability constraint \eqref{eq:C2}. For the  latency constraint \eqref{eq:C1}, the extension of Result \ref{res:equality} is less obvious, and the reasoning depends on the type of delay feedback model: 
\begin{itemize}
\item For $D(\vec{n}_m)=d, \forall m$, we can simply consider Problem \ref{prob} with blocklength $N'=N-\lceil Md \rceil$, where $\lceil \cdot \rceil$ stands for the ceiling operator, and no delay penalty. Therefore the latency constraint is equivalent to the following equality: 
\begin{equation}\label{eq:Dconst}
\sum_{m=1}^M n_m=N-\lceil Md \rceil.
\end{equation}
\item For $D(\vec{n}_m)=r\sum_{i=1}^m n_i, \forall m$, lemma \ref{lem:NbetterThanP} should be cautiously employed. Indeed, increasing the blocklength of the first fragment by one leads to an increase in the feedback delay at each fragment by $\lceil r\rceil$. After $M$ transmissions, the additional delay is at most $M\lceil r\rceil$. 
We know that the optimal solution lies in the following interval 
\begin{equation}\label{eq:Dlin}
N-M\lceil r\rceil \leq \sum_{m=1}^M \left(n_m+D(\vec{n}_m) \right) \leq  N,
\end{equation}
since the right-hand side (RHS) inequality in \eqref{eq:Dlin} ensures the latency constraint, and the left-hand side inequality in \eqref{eq:Dlin} is necessary for the optimal solution. Indeed, without this inequality, it is still possible to expand the first round by one without violating the latency constraint, hence obtaining a better solution than the optimal one, which leads to a contradiction. 
\end{itemize}

In addition to the previous result and lemmas, we have the following result, which only holds when $D(\vec{n}_m)=0, \forall m$. 
\begin{res}\label{prop:IncreaseM} 
When  $D(\vec{n}_m)=0, \forall m$, and given $T_{\rm rel}$ and $N$, increasing the number of retransmissions $M$ always yields a lower optimal average energy.
\end{res}
\begin{IEEEproof}
See Appendix \ref{anx:IncreaseM}.
\end{IEEEproof}
\smallskip

Result \ref{prop:IncreaseM} implies that when ideal feedback and no delay are guaranteed, a HARQ mechanism is always beneficial, i.e., it is always preferable to split the sub-codewords into smaller sub-codewords.

\subsection{Low Complexity Algorithm with Dynamic Programming}
In the previous subsection, the set of feasible points has been reduced without losing optimality (as established from Result \ref{res:equality}) and as a consequence, the search for the optimal solution of Problem \ref{prob} has been simplified. 
Nevertheless, due to the lack of convexity or other favorable properties for the objective function, an exhaustive search seems to be required. That involves the need for power quantization, which introduces an approximation error (denoted by $\theta$). 
The procedure is as follows: first, $\vec{n}_{M-1}$ and $\vec{P}_{M-1}$ are fixed; then, $n_M$ is obtained through \eqref{eq:C1} with equality, and $P_M$ is subsequence obtained through a bisection method for solving \eqref{eq:C2} with equality. The bisection method is possible since Lemma~\ref{res:epsbDecreasingP2} establishes the monotonicity of $\varepsilon_M$. Finally, it remains to perform a $2(M-1)$-D exhaustive search to solve Problem~\ref{prob}. The described brute force algorithm yields a complexity in $\mathcal{O}(N^{M-1}(1/\theta)^{M-1}\log(1/\theta))$. If $M$ is small enough (typically less than 3), the algorithm can be implemented. However, when $M$ is large, performing exhaustive search is prohibitively costly and an alternative approach is required. For that, we propose an algorithm based on dynamic programming (DP). We start from the case of zero delay feedback.

We assume the optimal solution to belong in $\mathcal{B}$ (as stated in Lemma \ref{lem:NbetterThanP}) so \eqref{eq:C1} and \eqref{eq:C2} become equalities. 
Let the state at the end of the round $m$
$$S_m=(N_{m}, V_{m},c_{m})$$
with  $N_{m}=\sum_{i=1}^{m} n_i$, $ V_{m}=\sum_{i=1}^{m} n_iP_i(P_i+2)/(P_i+1)^2$, and $c_{m}=Q^{-1}(\varepsilon_{m})$. The state sequence forms a Markov chain, i.e., $p(S_m|S_{m-1}, \cdots S_1)=p(S_m|S_{m-1})$ since we have 
\begin{eqnarray}
	N_m&=&N_{m-1}+n_m\\
	V_m&=&V_{m-1}+n_m\left(1-\frac{1}{(P_m+1)^2} \right)\label{eq:V}\\
	c_m&=&\frac{c_{m-1}\sqrt{V_{m-1}}+n_m\ln(1+P_m)}{\sqrt{V_m}}\label{eq:c} 
\end{eqnarray}
and the way to go from $S_{m-1}$ to $S_{m}$ depends only on the current round $m$ through $n_m$ and $P_m$. Notice that the assumption in Lemma \ref{lem:NbetterThanP} ensures $c_M =Q^{-1}(1- T_{\rm rel})$ and $0\leq c_1 \leq c_M \leq\sqrt{2B\ln2}/3$, while Result \ref{res:equality} ensures $N_M=N$.

The idea comes from the fact that the $m$ first components of the objective function can be written as follows 
\begin{equation}\label{eq:state}
	\sum_{i=1}^m n_iP_i\varepsilon_{i-1}=\sum_{i=1}^{m-1} n_iP_i\varepsilon_{i-1} +\Delta E(S_{m-1},S_m)
\end{equation}
where $\Delta E(S_{m-1},S_m)=n_mP_m\varepsilon_{m{-}1}$.
Let $E^\star(S_m)$ be the minimum average energy going to the state $S_{m}$. 
According to \eqref{eq:state}, it is easy to prove that 
\begin{align}\label{eq:viterbi}
\hspace{-0.1cm}E^\star(S_m){=}\underset{\forall\text{ possible }S_{m-1}}{\min}\lbrace \Delta E(S_{m-1},S_m)+E^\star(S_{m-1})\rbrace
\end{align}
since our problem boils down to the dynamic programming framework, and so Viterbi's algorithm can be used. 

Compared to the exhaustive search, the complexity is significantly reduced, but can be still very large  depending on the number of states $S_{m-1}$ and $S_{m}$ that has to be tested in \eqref{eq:viterbi}. 
First, we  see that the set of states $S_{m}$ for $m\in\{1, \cdots, M\}$ is not $\mathbb{R}^3$ but a much smaller set. 
Indeed the first component, we have $N_m\in\mathcal{N}_d=\lbrace 1,2,...,N\rbrace$. For the second component, we have $V_m\in\mathcal{V}_d=(0,\min(N_m,c_m{+}\sqrt{c_m^2{+}2B\ln2}))$ since $\sum_{i=1}^m n_i\ln(1+P_i)-B\ln2=c_m\sqrt{V_m}$ and  $\sum_{i=1}^m n_i\ln(1+P_i)\geq V_m/2$ (as $P(P+2)/(1+P)^2< 2\ln (1+P)$), which implies that $V_m/2-B\ln2 \leq c_m\sqrt{V_m}$ and so $V_m\leq c_m+\sqrt{c_m^2+2B\ln2}$. For the third component, we need the next Lemma 
\begin{lem}\label{lem:GeneralDecreasingEps}
	If $D(\vec{n}_m)=0$ then the optimal solution $(\vec{n}_M^\star,\vec{P}_M^\star)$ satisfies $\varepsilon_{1}>\varepsilon_2>...>\varepsilon_M$, and so $c_{1}<c_2<...<c_M$.
\end{lem}
\begin{IEEEproof}
	See Appendix \ref{anx:GeneralDecreasingEps}.
\end{IEEEproof}
According to Lemma \ref{lem:GeneralDecreasingEps}, we have $c_m\in\mathcal{C}_d= [0,Q^{-1}(1-T_{\rm rel})]$. 

Now focusing on the $S_{m-1}$ case, we straightforwardly have 
\begin{eqnarray}
	(m{-}1)n_{min}&\leq N_{m{-}1}\leq &N_m{-}n_{min}\label{ineq:Nm-1}\\
	V_m-n_m&\leq V_{m{-}1} \leq & \min\lbrace{V_m,N_{m-1}}\rbrace\label{ineq:Vm-1}
\end{eqnarray} 
where $n_{min}$ is the minimum blocklength of the transmitted packets. Finally, given the target $S_{m}$ and $(N_{m-1},V_{m-1})$ there is at most one feasible $c_{m-1}$ which emerges from \eqref{eq:V}-\eqref{eq:c}
\begin{equation}\label{eq:DP_findcm}
	c_{m{-}1}=\frac{ c_m\sqrt{V_m}{+}2(N_m{-}N_{m{-}1})\ln\big(1{-}\frac{V_m{-}V_{m{-}1}}{N_m{-}N_{m{-}1}}\big)}{\sqrt{V_{m{-}1}}}.
\end{equation}

Let us now focus on the initialization. When $M=1$, the states $S_1$ are 2D since given $(N_1,c_1)$ there can be only one feasible $P_1$ (and so $V_1$) which satisfies the equation $\varepsilon_1(N_1,P_1)=Q(c_1)$. Therefore we start from $M=2$. To find $E^\star(S_2)$, we need to minimize over only one variable ($N_1$), which renders this case computationally easier. Formally,
\begin{align}
	E^\star(N_2,V_2,c_2)&=\underset{N_1}{\min} \quad N_1P_1+n_2P_2\varepsilon_1(N_1,P_1)\label{eq:initializationDP}\\
	\quad\text{s.t.} \quad& n_2=N_2-N_1\nonumber\\
	& V_2=\frac{N_1P_1(P_1+2)}{(P_1+1)^2}+\frac{N_2P_2(P_2+2)}{(P_2+1)^2}\nonumber\\
	& \varepsilon_2(N_1,P_1,n_2,P_2)=Q(c_2).\nonumber  
\end{align}

Letting the approximation error due to quantization of $V$ and $c$ be $\theta_V$ and $\theta_c$, respectively, then the complexity is of order $\mathcal{O}(MN^2(\frac{1}{\theta_V})^2\frac{1}{\theta_c})$. In other words, the complexity of the dynamic programming algorithm is linear with respect to $M$, whereas the complexity of exhaustive search is exponential in $M$. 

Extension of the above algorithm to the case of non-zero delay is easy when $D(\vec{n}_m)=d$ since we can simply reconsider the problem as having available blocklength $N'=N-\lceil Md \rceil$ and no delay penalty. When $D(\vec{n}_m)=r\sum_{i=1}^m n_i$ more changes are required: first, $N_m$ now represents the available latency at the $m$-th round, second, an additional data structure $\rm{N_{net}}$ is needed which stores the number of symbols sent disregarding the delays, and third to find every $E^\star((N_m,V_m,c_m))$ an additional search within the states $(N,V_m,c_m), \forall N\in[N_m-m\lceil r\rceil, N_m-1]$ is employed.
 
\section{Asymptotic Regime}\label{sec:Asympt}
The minimum average energy for sending a fixed number of $B$ information bits is a decreasing function with respect to the latency $N$. Indeed, as seen in Problem \ref{prob}, the optimal solution for a given $N$ is a feasible solution of $(N+1)$ and so equal or worse than the optimal solution for the latency $(N+1)$. In following result, we prove that the optimal solution converges to an asymptotic value when $N\to\infty$. 
\begin{res}\label{lem:AsymptN} 
When $N\to\infty$, the minimum average energy of Problem \ref{prob} for fixed $M$ is given by
\begin{eqnarray}
E^\star_{as}(M,B) & = & \underset{(E_1,\cdots, E_M)}{\min} \quad  r(E_1, \cdots, E_M)\nonumber\\
&\text{s.t.} & \sum_{m=1}^M E_m= E_{\rm No-HARQ}^\infty \label{eq:Ec}
\end{eqnarray}  
with 
\begin{equation}
\hspace{-0.15cm}r(E_1,{\cdots}, E_M){=}E_1{+}\sum_{m=2}^M Q\left(\frac{\sum_{i=1}^{m-1}E_i{-}B\ln2}{\sqrt{2\sum_{i=1}^{m-1}E_i}}\right)E_m \label{eq:AsymptObj}
\end{equation}
and $E_{\rm No-HARQ}^\infty{=}\frac{(Q^{-1}(1{-}T_{\rm rel}))^2}{2}\left(1{+}\sqrt{1{+}\frac{2B\ln 2}{(Q^{-1}(1-T_{\rm rel}))^2}} \right)^2$. 
\end{res}
\begin{IEEEproof}
	See Appendix \ref{anx:AsymptN}.
\end{IEEEproof}
\smallskip

\begin{figure}[htb]
\centering
\includegraphics[width=0.86\columnwidth]{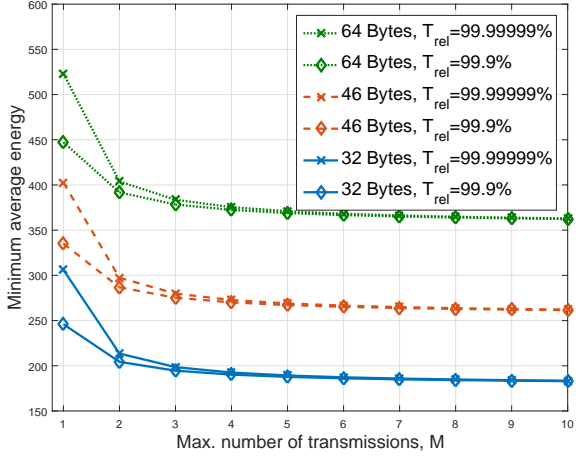}
\caption{Minimum average energy (when $N\to\infty$) versus $M$.}
\label{Fig:AsymptN}
\end{figure}

Note that $E_{\rm No-HARQ}^\infty$ corresponds to the required average energy when $N\to\infty$ for the case of no HARQ and can also be obtained from \cite[eq.(4.309)]{PolyanskiyThesis}. As an illustration, in Figure~\ref{Fig:AsymptN} we plot $E^\star_{as}(M)$ versus $M$ for different $B$ and $T_{\rm rel}$. 

When $N\to\infty$, a non-zero delay feedback - irrespectively of the model considered - does not impact the asymptote value since the latency constraint vanishes, which makes that Result~\ref{prop:IncreaseM} still holds. 

When $M$ also grows to infinity, we have an additional result.
\begin{res}\label{lem:AsymptNandM}
When $M\to\infty$, the asymptotic minimum average energy stated in Result \ref{lem:AsymptN} behaves as follows:
\begin{equation}
\lim_{M\to\infty} E_{as}^\star(M)=\int_0^{E_{\rm No-HARQ}^\infty}Q\left(\frac{E-B\ln2}{\sqrt{2E}}\right)\mathrm{d}E. \label{eq:AsymptNandM}
\end{equation}
\end{res}
\begin{IEEEproof}
	See Appendix \ref{anx:AsymptNandM}.
\end{IEEEproof}
\smallskip

Given $B$, increasing $T_{\rm rel}$ to $\overline{T}_{\rm rel}$ also increases $E_{\rm No-HARQ}^\infty$ to $\overline{E}_{\rm No-HARQ}^\infty$, which implies that $\lim_{M\to\infty}E_{as}^\star(M)< \lim_{M\to\infty} \overline{E}_{as}^\star(M)$.
In Figure~\ref{Fig:AsymptN}, these limit values cannot be distinguished and seem to coincide since they are very close to each other. This happens because,  as it easily can be shown,
$\lim_{M\to\infty} \overline{E}_{as}^\star(M){-}\lim_{M\to\infty}E_{as}^\star(M) < (1-T_{\rm rel})(\overline{E}_{\rm No-HARQ}^\infty-E_{\rm No-HARQ}^\infty)$
and $T_{\rm rel}$ is very small.

\section{Numerical Results and Discussion}\label{sec:num}
In this section, we provide numerical results to validate our analysis. We consider $B > 32$ bytes and $T_{\rm rel} > 99.99999\%$, i.e., $1-T_{\rm rel}\gg Q(\sqrt{2B\ln2}/3)\geq 1.7\cdot 10^{-10}$ always holds. Furthermore, we consider $n_1$ and $P_1$ such that $\varepsilon_1 <0.5$. Consequently, the assumption on $\mathcal{B}$ in Lemma \ref{lem:NbetterThanP} is not restrictive. The latency constraint \eqref{eq:C1} is simplified either according to \eqref{eq:Dconst} for fixed delay feedback model (including $D=0$) or according to \eqref{eq:Dlin} for the linear delay feedback model.

\begin{figure*}
\subfloat[Minimum average energy vs. $N$ for $B=32$ Bytes and $T_{\rm rel} =99.999\%$.]{\includegraphics[width=0.318\linewidth]{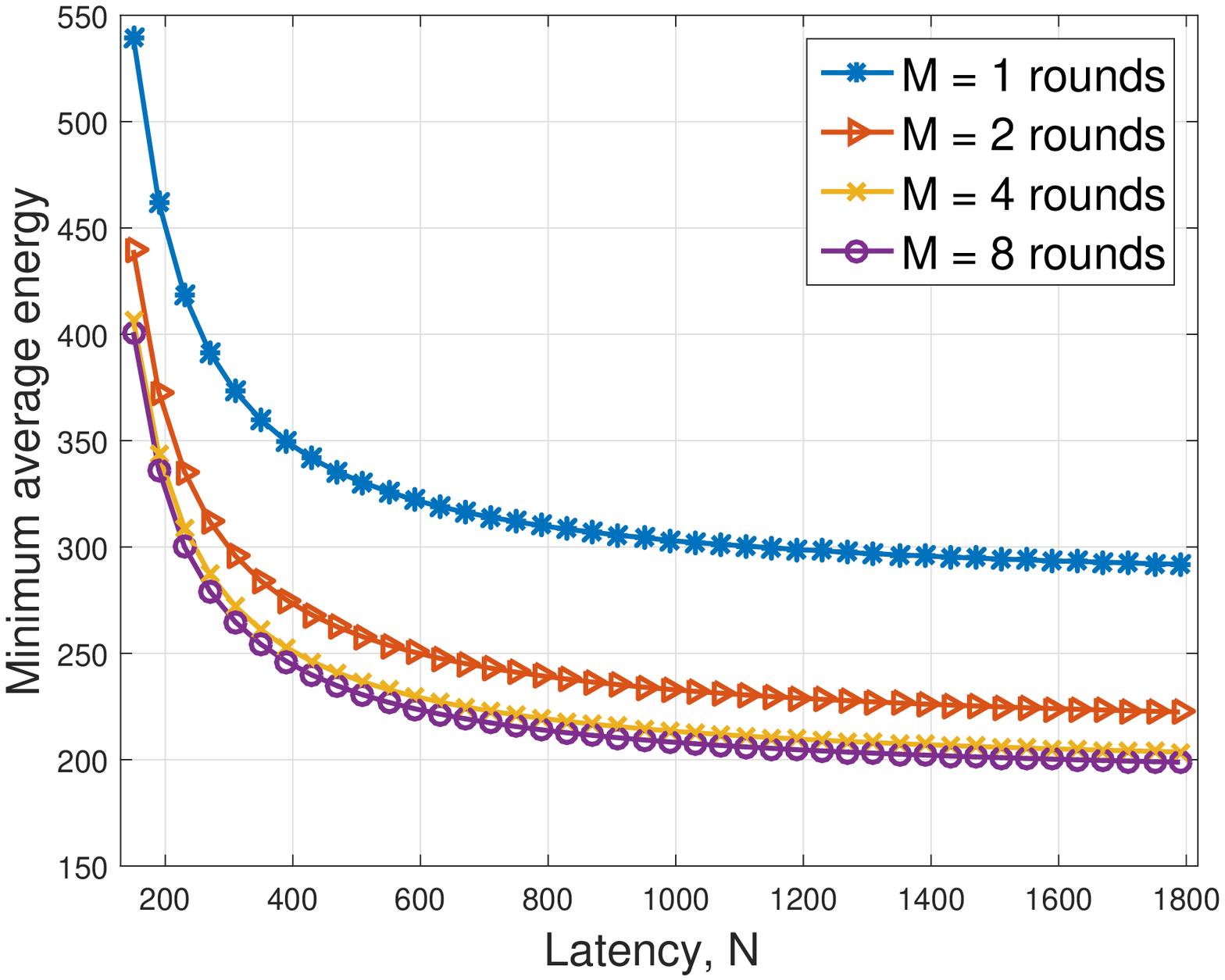}\label{Fig:EvsN}}
\hfil
\subfloat[Energy gain of $M$ rounds over no HARQ ($M{=}1$) vs. $N$ for $B=32$ Bytes and $T_{\rm rel} {=}99.999\%$.]{\includegraphics[width=0.333\linewidth]{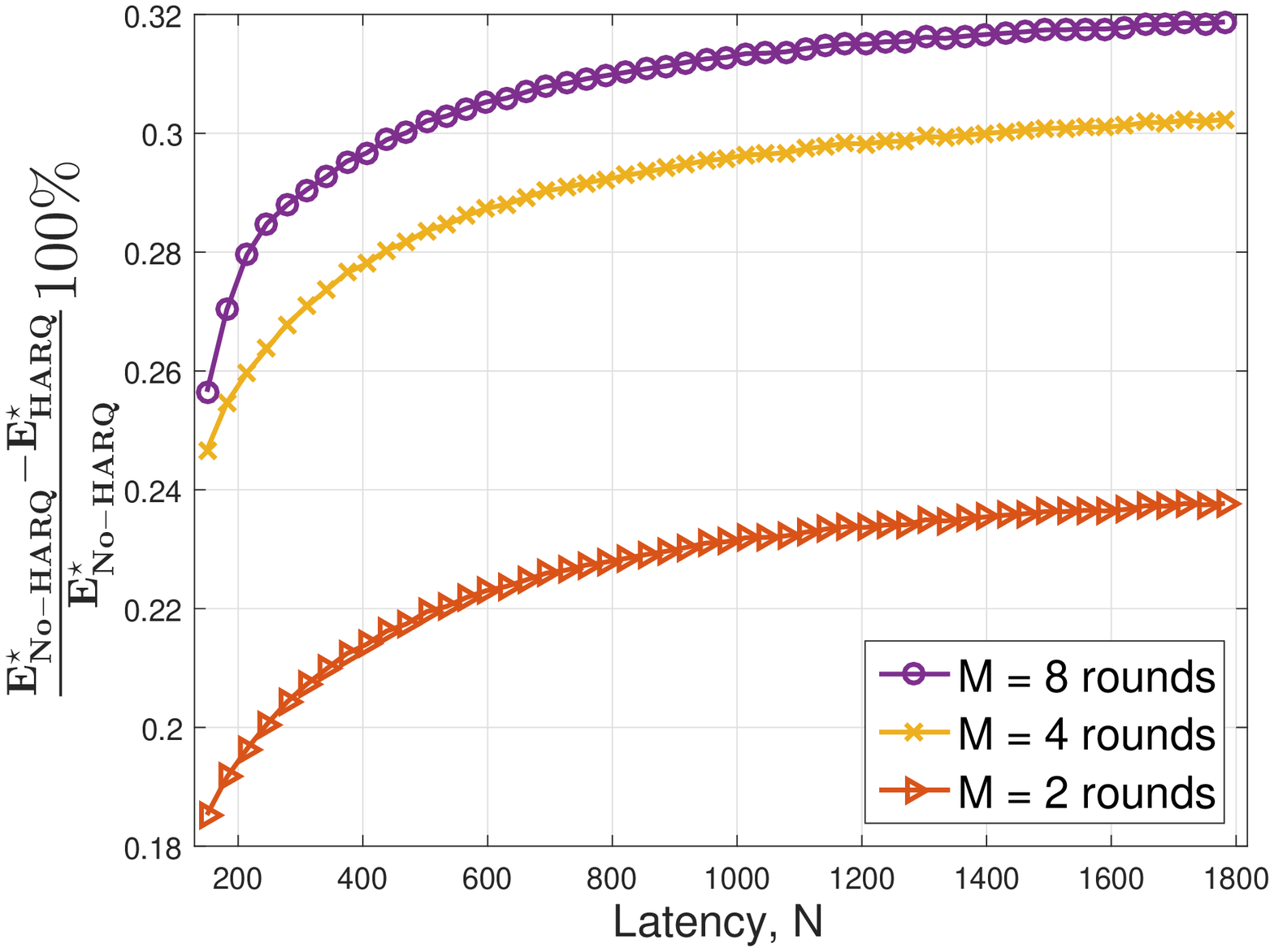}\label{Fig:GainVsN}}
\hfil
\subfloat[Energy gain vs. $B$ in the asymptotic regime ($N\to\infty$).]{\includegraphics[width=0.324\linewidth]{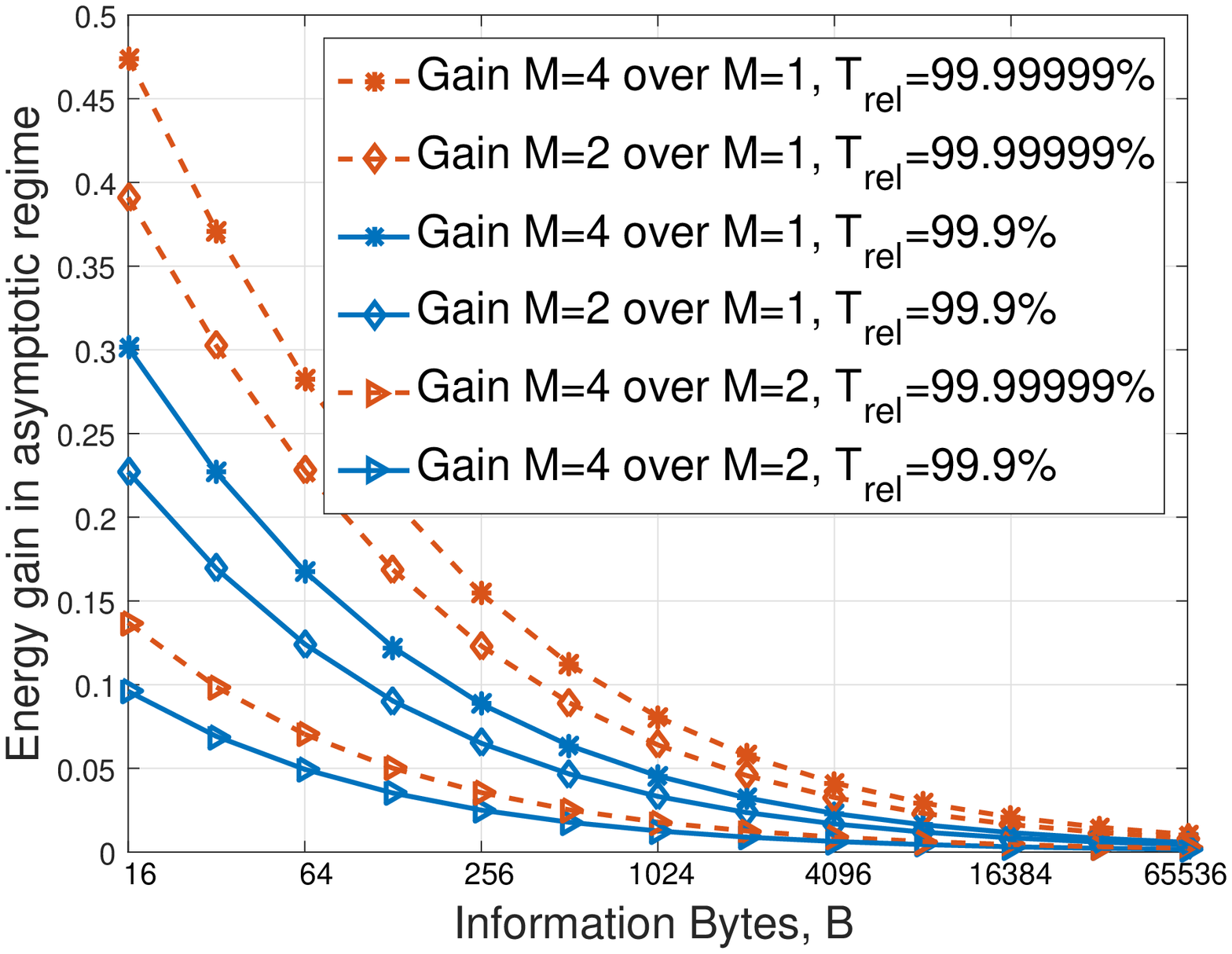}\label{Fig:Asympt_GainVsB}}
\caption{Performance analysis when $D(\vec{n}_m)=0$.}\label{Fig:PlotsZeroD}
\end{figure*}

First, we assume $D=0$. In Figure~\ref{Fig:EvsN}, we plot the minimum average energy versus $N$ and confirm that the energy for sending $B$ information bits decreases when $N$ increases. Additionally, the energy attains the asymptotic value predicted by Result \ref{lem:AsymptN}. Moreover, we confirm Result \ref{prop:IncreaseM}, since the minimum average energy decreases when $M$ increases for the case of zero delay feedback; however, the gain becomes negligible when $M$ is large enough. In Figure~\ref{Fig:GainVsN}, for the same $B$ and $T_{\rm rel}$ as in Figure~\ref{Fig:EvsN}, we plot the energy gain by using HARQ with $M$ rounds over $M=1$ (denoted as $E_{\rm No-HARQ}$). We observe that the energy gain monotonically increases when $N$ grows. As the latency constraint becomes more stringent, the benefit from employing HARQ diminishes. 
In Figure~\ref{Fig:Asympt_GainVsB}, we plot the energy gain for different values of $M$ versus $B$ when $N\to\infty$. The energies and the corresponding gains are derived using Result \ref{lem:AsymptN}. The higher the reliability or the lower $B$, the higher the gain. This remark also holds for non-zero delay feedback since we are in the asymptotic regime

\begin{figure}[htb]
\centering
\includegraphics[width=0.87\columnwidth]{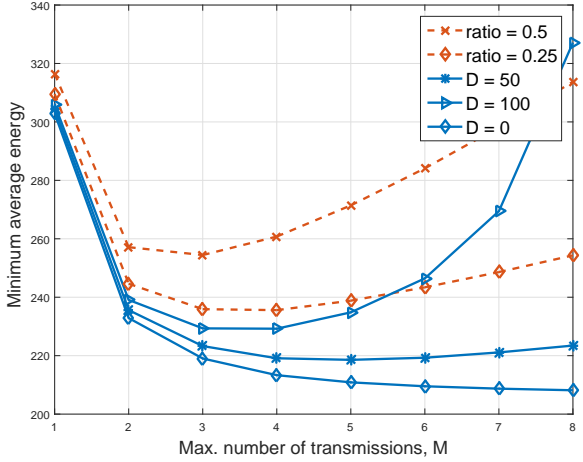}
\caption{Minimum average energy vs. $M$ for $B{=}32$ Bytes and $T_{\rm rel} {=}99.999\%$. }
\label{Fig:EVsM}
\end{figure}

We consider now $D(\vec{n}_M)\neq 0$ unless otherwise stated. 
In Figure~\ref{Fig:EVsM}, we plot the minimum average energy versus $M$ for different delay feedback models (solid lines for fixed delay and dashed line for the linear delay model). When $d>0$, splitting the packet/transmission in rounds is not always advantageous and we observe that an optimal bounded value of $M$, denoted by $M^\star$, exists. Indeed, for small values of $M$, the delay penalty is small and it is still of interest to split, whereas when $M$ is large, the value of $N'$ in \eqref{eq:Dconst} becomes very low and there is no gain to split further. The same statement holds when the linear delay feedback model is applied.

\begin{figure}[htb]
\centering
\includegraphics[width=0.87\columnwidth]{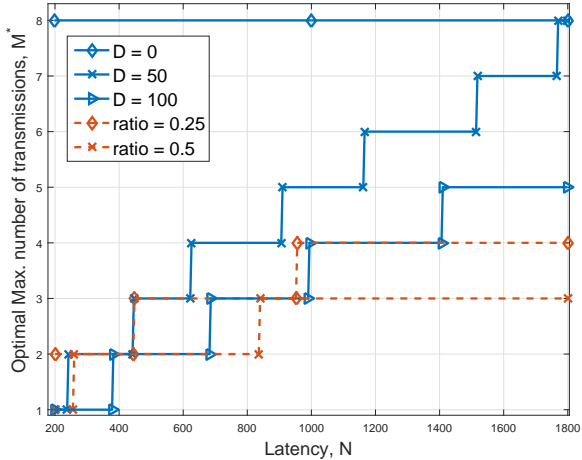}
\caption{$M^\star$ (assuming $M\leq 8$) vs. $N$ for $B{=}32$ Bytes and $T_{\rm rel} {=}99.999\%$.}
\label{Fig:MstarVsN}
\end{figure}

In Figure~\ref{Fig:MstarVsN}, we plot $M^\star$ versus $N$ restricting $M\leq 8$. The delay penalties become more significant when $N$ decreases when eventually prevents from using an HARQ mechanism. Therefore, $M^\star$ increases with respect to $N$. In the case of linear delay feedback model, $M^\star$ increases much slower than in the fixed delay feedback model since the effect of delay in the energy consumption is higher when $M$ increases. 

\section{Conclusion}\label{sec:ccl}
In this paper, we have characterized the energy-latency tradeoff in URLLC systems with retransmissions in the finite blocklength regime and showed how IR-HARQ can be optimized by tuning the number of rounds, the blocklength and the transmit power. A dynamic programming algorithm for solving the non-convex average energy minimization problem subject to URLLC constraints is provided. The main takeaway of this paper is that a properly optimized IR-HARQ scheme can be beneficial in terms of energy as long as the feedback delay is reasonable compared to the packet size. 
Future work could study how frequency or space diversity can alter the tradeoff and the IR-HARQ design. Further extensions of this framework may include the analysis of fading and multi-antenna systems with both perfect and imperfect channel knowledge.

\appendices

\section{Proof of Lemma~\ref{res:epsbDecreasingP2}}\label{anx:res1}
Let us denote by $\partial \varepsilon_M/\partial P$ the derivative function of  $P\mapsto \varepsilon_M(\vec{n}_M^\dag, \vec{P}_{M-1}^\dag, P)$. We will prove that   $\partial \varepsilon_M/\partial P_{|P=P_M^\dag} <0$. By change of variables $y=1/(P+1)^2$ and putting  $y^\dag=1/(P_M^\dag+1)^2$, we show that 
\begin{eqnarray}
\nonumber \frac{\partial \varepsilon_M }{\partial P}< 0 \ \text{at} \ P=P_M^\dag &\Leftrightarrow& \frac{\partial \varepsilon_M}{\partial y}> 0 \ \text{at} \ y=y^\dag\\ &\Leftrightarrow& h(y)>0 \ \text{at} \ y=y^\dag \label{eq:monotonicity_PM}
\end{eqnarray}
where $h(y)=k_2-yk_1+n_M(1-y+y\ln(y)/2)$ with $k_1=\sum_{i=1}^{M-1} n_i\ln(1{+}P_i)-B\ln2$ and $k_2=\sum_{i=1}^{M-1} n_i\left(1-1/(1+P_i)^2\right)$. It is easy to prove that $h(y)$ is a monotonically decreasing function. If $h(1)\geq  0$, then \eqref{eq:monotonicity_PM} is straightforwardly satisfied. If $h(1)<0$, then it exists $y_0\in(0, 1)$ such that $h(y_0)=0$. So for $y\in [y_0,1]$, we get $h(y)\leq 0$, which implies that $\varepsilon_M$ is decreasing with respect to $y$. As a consequence, for $y\in [y_0,1]$ and so the corresponding $P(y)$, we have $\varepsilon_M(\vec{n}_M^\dag, \vec{P}_{M-1}^\dag, P(y)) \geq \varepsilon_M(\vec{n}_M^\dag, \vec{P}_{M-1}^\dag, 0)= \varepsilon_{M-1}(\vec{n}_{M-1}^\dag, \vec{P}_{M-1}^\dag)$ which prevents to have $P(y)= P_M^\dag$ according to the assumption $\varepsilon_{M-1}\geq \varepsilon_{M}$ on the analyzed point. Consequently, $y^\dag$ does not belong to $[y_0,1]$, and belongs to $(0,y_0)$ where \eqref{eq:monotonicity_PM} holds.

\section{Proof of Lemma~\ref{lem:NbetterThanP}}\label{anx:res2}
Let $\varepsilon_1=Q(F_1(a))$ and $\varepsilon_M=Q(F(a))$ where
$$F_1(a)=\frac{g_1(a)-c}{\sqrt{g_2(a)}} \ \text{and} \ F(a)=\frac{g_1(a)+c_1-c}{\sqrt{g_2(a)+c_2}},$$
with $g_1(a)=an_1\ln(1+\frac{P_1}{a})$, $g_2(a)=an_1(1-1/(1+P_1/a)^2)$, $c_1=\sum_{m=2}^M n_m\ln(1+P_m)$,  $c_2=\sum_{m=2}^M n_m(1-1/(1+P_m)^2)$, and $c=B\ln2$.
As we consider a point in $\mathcal{B}$, we get   
\begin{align}
 \varepsilon_1<0.5 \Leftrightarrow an_1\ln(1+P_1/a)>c \Rightarrow E_1>B\ln2 \label{eq:eps1Lower0.5}
\end{align}
where $E_1=n_1P_1$. To prove \eqref{eq:eps1Lower0.5}, we use the inequality $\ln(1+x)\leq x \text{ when } x\geq 0$. Once again, belonging to $\mathcal{B}$ leads to 
\begin{align}
F_1(a)\leq F(a)\leq \sqrt{2B\ln2}/3.  \label{eq:F1andFlessR0}
\end{align}

We want to show that $\varepsilon_1$ and $\varepsilon_M$ are decreasing functions with respect to $a$, i.e., $F_1'(a)\geq 0$ and $F'(a)\geq 0$ where $f'(a)$ stands for ${\rm d}f/{\rm d}a$ for any mapping $f$.
As $g_1(a),g_2(a),g_1'(a)$ and $g_2'(a)$ are strictly positive, we have  
\begin{eqnarray}
F_1'(a)\geq 0 & \Leftrightarrow & 2g_1'(a)g_2(a)\geq g_2'(a)(g_1(a)-c)\label{eq:F1increasing}\\
& \Leftrightarrow & c\geq E_1H(P_1/a)\label{eq:AnalysisF1increasing}
\end{eqnarray}
and
\begin{align}
\hspace{-4mm}F'(a)\geq 0 \Leftrightarrow & 2g_1'(a)(g_2(a)+c_2)\geq g_2'(a)(g_1(a)+c_1-c)\label{eq:Fincreasing}\\
{\Leftrightarrow}& c\geq E_1H(P_1/a)+(c_1-K(P_1/a)c_2)\label{eq:AnalysisFincreasing}
\end{align}
where
$$x\mapsto H(x)=\frac{2x+4-\ln(1+x)(\frac{4}{x}+x+3)}{x(x+3)},$$
and 
$$x\mapsto K(x)=\frac{2(x+1)^3\left(\ln(1+x)-\frac{x}{x+1}\right)}{x^2(x+3)}.$$
After some algebraic manipulations, \eqref{eq:F1increasing} and \eqref{eq:Fincreasing} are equivalent to 
\begin{eqnarray}
F_1(a)& \hspace{-2.5mm}\leq&\hspace{-2.5mm}\frac{2g_1'(a)\sqrt{g_2(a)}}{g_2'(a)}=\sqrt{E_1}W(P_1/a,0)\label{eq:eps1Condition}\\
F(a) &\hspace{-2.5mm}\leq&\hspace{-2.5mm}\frac{2g_1'(a)\sqrt{g_2(a)+c_2}}{g_2'(a)}=\sqrt{E_1}W(P_1/a,\frac{c_2}{E_1})\label{eq:epsCondition}
\end{eqnarray}
with \vspace{-5mm}
\begin{equation}\label{eq:W}
  (x,y)\mapsto W(x,y)=K(x)\sqrt{y+\frac{x+2}{(1+x)^2}}.
\end{equation}
We want now to prove that either \eqref{eq:AnalysisF1increasing} or \eqref{eq:eps1Condition} holds for any $x>0$, and either \eqref{eq:AnalysisFincreasing} or \eqref{eq:epsCondition} holds for any $x>0$. For that, we split the analysis into two intervals on $x$.
\begin{itemize}
\item{If $x\in(0,484)$:} the function $x\mapsto W(x,0)$ is a positive unimodal function converging to zero when $x \to \infty$. For $x\in(0,484)$, it is easy to check that $W(x,0)\geq W(0,0) = \sqrt{2}/3$. As $W(x,y)>W(x,0)$
 for any $y\geq 0$, we obtain that $\sqrt{E_1}W(x,y)\geq\sqrt{E_1}W(x,0) \geq \sqrt{2E_1}/3$. Due to \eqref{eq:eps1Lower0.5}, we have $\sqrt{E_1}W(x,y)\geq\sqrt{E_1}W(x,0) \geq \sqrt{2B\ln2}/3$.  According to \eqref{eq:F1andFlessR0}, we check that  $\sqrt{E_1}W(x,y)\geq\sqrt{E_1}W(x,0) \geq F(a)\geq F_1(a)$. Therefore, \eqref{eq:eps1Condition} and \eqref{eq:epsCondition} hold.
\item{If $x\in [484,\infty)$:} in that interval, we can see that $H(x)\leq 0$, which implies that \eqref{eq:AnalysisF1increasing} holds.

It now remains to check that either \eqref{eq:AnalysisFincreasing} or \eqref{eq:epsCondition} holds. For doing so, we distinguish two cases: 
\begin{itemize}
\item[$\circ$] If $c_1 \leq 10.37c_2$:  one can check that $K(x)$ is an increasing function. Therefore for $x \geq 484$, we get $K(x)\geq K(484)>10.37$. Consequently, $c_1-K(x)c_2<0$. As $H(x)\leq 0$ too for $x\geq 484$, it is easy to show that \eqref{eq:AnalysisFincreasing} holds.
\item[$\circ$] If $c_1> 10.37c_2$: this inequality leads to 
$$\sum_{m=2}^M n_m\ln(1+P_m){-}10.37n_m\left(1-\frac{1}{(1+P_m)^2}\right)>0 $$
which forces that there exists at least one $m_x\in\lbrace 2, ...,M\rbrace$ such that $n_{m_x}\ln(1+P_{m_x})>10.37n_{m_x}(1-1/(1+P_{m_x})^2)>0\Rightarrow P_{m_x}>31866$ which implies that $c_2\approx \sum_{m\in \lbrace 2,..,M\rbrace\setminus m_x}n_m(1-1/(1+P_m)^2)+n_{m_x}\Rightarrow c_2>n_{m_x}$. Consequently, according to \eqref{eq:W}, $\sqrt{E_1}W(x,c_2/E_1)\geq K(484)\sqrt{n_{i_x}}\geq 10.37\cdot\sqrt{1}$. If \eqref{eq:epsCondition} does not hold, one can see that $\varepsilon_M<Q(10.37)\approx 1.7{\cdot}10^{-25}$. As this error does not correspond to any reasonable operating point, we consider that \eqref{eq:epsCondition} holds. 
\end{itemize}
\end{itemize}

\section{Proof of Result~\ref{prop:IncreaseM}}\label{anx:IncreaseM}
Consider the last round $M$ where for the optimal point $(\vec{n}_M^\star,\vec{P}_M^\star)$, we know that $\varepsilon_{M-1}>  \varepsilon_M$ (see Lemma \ref{lem:1} and its related proof for more details). For $x\in[0,n_M^{\star}]$, let 
$$F(x)=Q\left(\frac{x\ln(1+P_M^{\star})+\sum_{i=1}^{M-1} n_i^{\star}\ln(1+P_i^{\star})-B\ln2}{\sqrt{x\frac{P_M^{\star}(P_M^{\star}+2)}{(P_M^{\star}+1)^2}+\sum_{i=1}^{M-1}n_i\frac{P_i^{\star}(P_i^{\star}+2)}{(P_i^{\star}+1)^2}}}\right).$$
We know that $F(0)=\varepsilon_{M-1}>\varepsilon_{M}=F(n_M^{\star})$ and that $F(\cdot)$ is a continuous (not necessary monotonically decreasing) function. Therefore, it exists $x_0\in(0,n_M^{\star})$ such that $F(0)<F(x_0)<F(n_M^{\star})$. If $F$ is smooth enough, it exists an integer  $\overline{n} \in \lbrace 1,2,...,n_M^{\star}-1 \rbrace$ (typically equal to $\lfloor x_0 \rfloor$  or $\lceil x_0 \rceil$) such that $\varepsilon_{M-1}>F(\overline{n})>\varepsilon_M$. Then, the new point of $M+1$ rounds, which is $(\vec{n}_{M-1}^\star,\overline{n}, n_M^\star-\overline{n},\vec{P}_{M-1}^\star,P_M^\star, P_M^\star)$, leads to the following average energy 
$$ \sum_{m=1}^{M-1} n_{m}^\star P_{m}^\star\varepsilon_{m-1}+ \overline{n} P_{M}^\star  F(\overline{n})+ (n_M^\star-\overline{n}) P_{M}^\star  \varepsilon_{M-1}, $$  
which is smaller that the average energy provided by the point $(\vec{n}_M^\star,\vec{P}_M^\star)$. Obviously the reliability constraint (given by $\varepsilon_M$) remains unaltered and the latency constraint does not change since $D(\vec{n}_m)=0$. So increasing the number of transmissions to $M+1$ improves the optimal operating point of $M$ transmissions.

\section{Proof of Lemma~\ref{lem:GeneralDecreasingEps}}\label{anx:GeneralDecreasingEps}
To prove the lemma, we will prove that if for some solution the states $\tilde{S}_{i-1}$, $\tilde{S}_i$,$\tilde{S}_{i+1}$ satisfy $\tilde{c}_{i-1}\geq \tilde{c}_i$ and $\tilde{c}_i<\tilde{c}_{i+1}$, then there exists a better solution, thus it cannot be the optimal one. Therefore, if for the optimal solution for some $i$ we know $c_{i+1}^\star>c_i^\star$ then it must $c_{i}^\star>c_{i-1}^\star$ and since from Lemma \ref{lem:1} we know $c_M^\star>c_{M-1}^\star$, this Lemma is proved by induction. 

To prove the existence of a better solution we only have to prove the superiority of a configuration of $M-1$ rounds that goes directly from the state $\tilde{S}_{i-1}$ to $\tilde{S}_{i+1}$ using one fragment of blocklength $n_i+n_{i+1}$  and has exactly the same configuration before and after those states (then due to proposition \ref{prop:IncreaseM} there exists an even better configuration with $M$ rounds) Hence, we only need to prove:
\begin{equation}
\Delta E(\tilde{S}_{i-1},\tilde{S}_{i})+\Delta E(\tilde{S}_{i},\tilde{S}_{i+1})\geq\Delta E(\tilde{S}_{i-1},\tilde{S}_{i+1}).\label{ineq:LessDE}
\end{equation}
Since a zero delay penalty is assumed, using \eqref{eq:C1} \& \eqref{eq:C2} with equalities allow to derive that
\begin{equation*}
\Delta E(S_{k-1},S_k)=n_kP_k\varepsilon_{k-1}=n_k(e^{\frac{\gamma_k}{n_k}}-1)Q(c_{k-1})
\end{equation*}
where $\gamma_k =c_k\sqrt{V_k}-c_{k-1}\sqrt{V_{k-1}}>0$. Since $\tilde{c}_{i-1}\geq \tilde{c}_i$, to prove \eqref{ineq:LessDE} it suffices to prove that
\begin{equation*}
\tilde{n}_ie^{\frac{\tilde{\gamma}_i}{\tilde{n}_i}}+\tilde{n}_{i+1}e^{\frac{\tilde{\gamma}_{i+1}}{\tilde{n}_{i+1}}}\geq(\tilde{n}_i+\tilde{n}_{i+1})e^{\frac{\tilde{\gamma}_i+\tilde{\gamma}_{i+1}}{\tilde{n}_i+\tilde{n}_{i+1}}}.
\end{equation*}
Changing variables as $\lambda_l=\frac{\tilde{n}_l}{\tilde{n}_i+\tilde{n}_{i+1}}$ and $x_l=\frac{\tilde{\gamma}_l}{\tilde{n}_l}$, $ l{\in}\lbrace i{,}i{+}1 \rbrace$:
\begin{equation*}
\lambda_i e^{x_i}+\lambda_{i+1} e^{x_{i+1}}\geq e^{\lambda_i x_i+\lambda_{i+1} x_{i+1}},
\end{equation*}
which holds due to the convexity of the exponential function.

\section{Proof of Result~\ref{lem:AsymptN}}\label{anx:AsymptN}
We consider $E_i= n_i^\star P_i^\star$ where $n_i^\star$ and $P_i^\star$ are the $i$-th blocklength and power components of $(\vec{n}_M^\star,\vec{P}_M^\star)$ respectively. Notice that each $E_i$ depends on $N$.
Let us assume that it exists at least $i_0\in\lbrace 1,2,...,M \rbrace$ such that $\Lim{N}{\infty}E_{i_0}=\infty$. According to Lemma \ref{lem:GeneralDecreasingEps}, we know that $\varepsilon_1> \varepsilon_2>...\varepsilon_M=1-T_{\rm rel} >0$ at the optimal point. Consequently  the minimum average energy $E_1+\sum_{i=2}^M\varepsilon_{i-1}E_i\to\infty$ too. For at least one finite $N$, say $N_f$, the optimal point leads to a finite minimum average energy. For any $N > N_f$, the optimal solution cannot increase the minimum average energy since the optimal solution at $N_f$ is a feasible point of Problem \ref{prob} for $N$. So the minimum average energy is upper bounded when $N\to\infty$. Therefore, $\Lim{N}{\infty}E_i<\infty ,\forall i\in\lbrace 1,2,...,M \rbrace$. 

When $N\to\infty$, we know that the delay feedback model does not have an impact on the latency constraint, so we can apply the results obtained for $D=0$.  
According to Lemma \ref{lem:NbetterThanP}, we also know that it is preferable to increase the blocklength rather than the power in order to save energy. Therefore, when $N\to\infty$, we have to take $n_1^\star$ as large as possible, i.e., $\Lim{N}{\infty}n_1^\star=\infty$. Similar arguments can be applied to the other rounds, i.e., $\Lim{N}{\infty}n_i^\star=\infty$ with $i\in\{2, \cdots, M \}$. As $\Lim{N}{\infty}E_i<\infty$, we get $\Lim{N}{\infty}P_i^\star{=}0$. 
By using $N\to\infty$ in 
$$\frac{P_i^\star}{P_i^\star+1}\leq \ln(1+P_i^\star) \leq P_i^\star$$
and the fact that  $\Lim{N}{\infty}P_i^\star{=}0$, we easily obtain that $E_i=\Lim{N}{\infty}n_i^\star \ln(1+P_i^\star)$. Plugging this previous equation in

 \eqref{eq:epsb} leads to  
\begin{align}
\Lim{N}{\infty} \varepsilon_m = Q\bigg(&\frac{\sum_{i=1}^m E_i-B\ln2}{\sqrt{2\sum_{i=1}^m E_i }}\bigg).\label{eq:epsAsymptN}
\end{align}
Putting $m=M$ in \eqref{eq:epsAsymptN} and using \eqref{eq:C2} with equality, we have 
\begin{equation}\label{eq:RelConstrAsympt}
\sum_{i=1}^M E_i=\frac{(Q^{-1}(1{-}T_{\rm rel}))^2}{2}\Big(1{+}\sqrt{1{+}\frac{2B\ln 2}{(Q^{-1}(1{-}T_{\rm rel}))^2}} {\Big)^2}
\end{equation}
where its RHS corresponds to the energy when no HARQ ($M=1$) is used and is denoted by $E_{\rm No-HARQ}^\infty$.

\section{Proof of Result~\ref{lem:AsymptNandM}}\label{anx:AsymptNandM}
\begin{figure}[htb]
	\centering
	\includegraphics[width=\columnwidth , height= 3.3cm ,trim={3cm 0 1.5cm 0},clip]{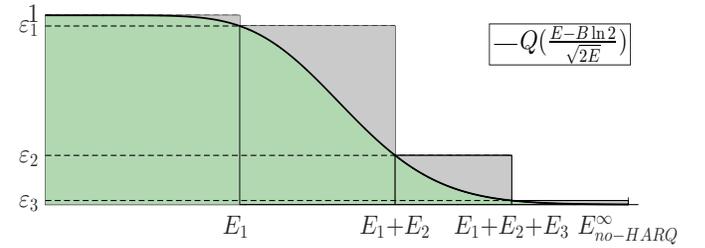}
	\caption{Geometrical interpretation of Result \ref{lem:AsymptN} for $M=4$.}
	\label{Fig:GeometricDescr}
\end{figure}

The function $E\mapsto Q((E-B\ln 2)/\sqrt{2E})$ is plotted in Figure \ref{Fig:GeometricDescr}. We also draw the $m$-th component of the objective function \eqref{eq:AsymptObj} of Result \ref{lem:AsymptN}, which corresponds to the area of the partially grey partially green rectangular box located from $\sum_{i=1}^{m-1} E_i $ to $\sum_{i=1}^{m} E_i$ with a level $\varepsilon_{m-1}$ (see \eqref{eq:epsAsymptN}). According to \eqref{eq:Ec}, the final point is $E_{\rm No-HARQ}^\infty$. 
Consequently, the sum of the green and the grey areas gives the value of the objective function \eqref{eq:AsymptObj}. It is evident that the function $E\mapsto Q((E-B\ln 2)/\sqrt{2E})$ coincides at the upper left corner of each rectangular box and is always inside each rectangular box (due to its decreasing monotonicity). Therefore, the value of the objective function \eqref{eq:AsymptObj} cannot be lower than the green area. When $M\to\infty$, we can decrease the width of each rectangular box converging to a solution that includes only the green area. Consequently, the minimum energy spent converges to the green area, which is identical to the Riemann integral of $E\mapsto Q((E-B\ln 2)/\sqrt{2E})$ from $0$ to $E_{\rm No-HARQ}^\infty$.

\bibliographystyle{IEEEtran}
\bibliography{LibraryICC2018}

\end{document}